\documentclass[aps,prl,twocolumn,showpacs,groupedaddress]{revtex4}
\usepackage{graphicx}

\begin{document}

\author{X. X. Yi$^1$ , H. J. Wang $^2$, H. T. Cui$^1$ and C. M. Zhang$^1$ }
\affiliation{$^1$ Institute of Theoretical Physics, Northeast
Normal University,
Changchun 130024, China\\
$^2$ Department of Chemistry, Hebei University, Baoding 017002,
China\\}
\title{Two-component Fermi gas in a Harmonic Trap}

\begin{abstract}
We consider a mixture of two-component Fermi gases at low
temperature. The density profile of this degenerate Fermi gas is
calculated under the semiclassical approximation. The results show
that the fermion-fermion interactions make a large correction to
the density profile at low  temperature. The phase separation of
such a mixture is also discussed for
 both attractive and repulsive interatomic
interactions, and the numerical calculations demonstrate the exist
of a stable temperature region $T_{c1}<T<T_{c2}$ for the mixture.
In addition, we give the critical temperature of the BCS-type
transition in this system beyond the semiclassical approximation.
\end{abstract}

 \maketitle

\section{ Introduction}
Since the  experimental realization of Bose-Einstein condensation
in dilute gases of rubidium[1-4], sodium[5,6], lithium[7], and
hydrogen[8], a great deal of interest in trapped ultra-cold atoms
has concentrated on the topic of trapped degenerate Fermi gas.
However, it is difficult to achieve a  degenerate state for
fermonic atoms,  because the $s-$wave collisions between fermions
in a same state are suppressed by the Pauli principle, and the
$p-$wave scattering as well as the dipole-dipole magnetic
interaction are very weak at low temperature. The successful
demonstration of overlapping condensates in different spin states
of rubidium [9,10] and sodium [11] open a door to study the
degenerate fermionic gas, because we can cool down one component
of a mixture by sympathetic cooling. Inspired by this observation
a number of experiments has been conducted on systems of
Bose-Fermi mixtures, and most recently, using two-component
evaporative cooling strategy, DeMarco and Jin[12] have succeeded
in cooling fermonic atom gas down to about 0.5 $T_F$(300$nK$,
depends on the trapping frequencies and the number of the trapped
atoms). Below this temperature  quantum degeneracy behaves as a
barrier to evaporative cooling and as a modification of the
classical thermodynamics. For the experiment reported by Demarco
and Jin[12], the atom $^{40}K$ were trapped in two magnetic
sublevels, $|F=9/2, m_F=9/2\rangle $ and $|9/2,7/2\rangle$, this
mixture of atoms  states is metastable against $m_F$ which changes
collisions at low temperature, therefore the atom in each state is
separately conserved. In such a two-component mixture of trapped
spin-polarized $^{40}K$ atoms, interactions between atoms in
different hyperfine states are much larger than those among atoms
in the same state. Indeed, under this approximations a relatively
high  temperature $T_c$ for a BCS-type phase transition was
predicted[13-14].

The purpose of this paper is to examine the properties including
normal and BCS-type phase transition of such trapped atoms. The
normal state properties of such  a system were studied under the
semiclassical approximation in Ref.[15-18]. However, as the
interactions between $^{40}K$ atoms in the two different hyperfine
states are considerable strong, it is important to include the
effect of these interactions in any realistic treatment of the
system. The present paper extends the analysis of Ref.[13-18] by
considering both the discrete nature of the trapped atom and the
effects of the interactions among them.

The paper is organized as follows. In Sec.II, we analyze the
influence of the trap potential and the interactions on the
density profile of the trapped atomic gas. We show that the cloud
of the trapped atoms is compressed (diluted) for the case of
attractive(repulsive) interactions. The stability properties of
the trapped two-component Fermi gases are considered in Sec.III,
and in Sec.IV we investigate the BCS-type transition in the system
by taking the discrete nature into account. We find that the
discrete nature is indeed make sense, they decrease the BCS-type
transition temperature. The results for the BCS-type transition
are beyond the semiclassical
approximation. Finally, we summarize our results in Sec. V.\\
\section {Density profile of a trapped interacting two-component Fermi gas}

We consider a dilute gas which consists of interacting fermionic
two-level atoms trapped in an external potential $V_0(r)$. As the
gas is dilute, the interactions mainly happen through two-body
collisions. Furthermore, because the $s-$wave scattering length
between fermions in a state is suppressed, and the p-wave
scattering is greatly reduced due to the presence of the
centrifugal barrier,   we may neglect the interactions between
fermions in the same hyperfine state and only consider the s-wave
scattering between the fermions in different hyperfine states.
Under this consideration, the system is then described by the
following Hamiltonian
\begin{eqnarray}
H&=&H_1+H_2+H_{int},\nonumber\\
H_i&=&\int dr \psi_i^{\dagger}(\frac{p^2}{2m}-\mu +V_0(r))\psi_i(r), (i=1,2)\nonumber\\
H_{int}&=&v_0\int dr dr^{'} \psi_1^{\dagger}(r)\psi_2^{\dagger}(r^{'})
\delta(r-r^{'})\psi_2(r^{'})\psi_1(r),
\end{eqnarray}
where $m$ is the mass of  one fermion, and the interatomic
potential has been approximated by a constant potential
$v_0\delta(r-r^{'})$. $\psi_i(r)$  stands for the annihilation
operator of a fermion at position $r$ in the hyperfine state
$|i\rangle$, and it obeys
 the usual fermionic anticommutation rules. The trapping potential is for simplicity
taken  to be   an isotropic harmonic oscillator potential
$V_0(r)=\frac 1 2 m\omega^2r^2$, and the trapping frequency is the
same for each hyperfine state. In addition to what we stated
above, we have assumed that the number of particles $N$ in each
state is the same such that we only have one chemical potential
$\mu$. As the critical temperature for a BCS-type transition is
maximum when the number of particles in the two hyperfine state is
equal[13,14], so this configuration  has most experimental
relevance. Indeed, in the experiment reported in Ref.[12], the
number of the atoms in state $m_F=9/2$ and $m_F=7/2$  are
approximately equal. The noninteracting case is achieved by
setting $v_0=0$; this limit has been discussed in ref.[15,16,18]
 for one component Fermi
gas within the semiclassical approximation. In this section, we
are interested in the effect of the interaction and the trapping
potential on the density profile under the semiclassical
approximation. We can, therefore, ignore any pairing correlations
leading to BCS-type transition, and use the mean field Hamiltonian
\begin{eqnarray}
H_{m1}&=&(\frac{p^2}{2m}-\mu+\frac 1 2
m\omega^2r^2)+v_0\langle\psi_2^{\dagger}(r)\psi_2(r)\rangle,
\nonumber\\ H_{m2}&=&(\frac{p^2}{2m}-\mu+\frac 1 2
m\omega^2r^2)+v_0\langle\psi_1^{\dagger}(r)\psi_1(r)\rangle,
\end{eqnarray}
this equation comes from eq.(1) straightforwardly. Here, $H_{mi}$
describes the effective Hamiltonian for component $i$. $\langle
\psi_i^{\dagger}(r)\psi_i(r)\rangle$ is the standard Hartree-Fock
result for a hard sphere
 interaction model.
In order to get some analytical results, we study the  density
profile here within the semiclassical approximation, it is given
from eq.(2) that
\begin{equation}
n_i(r)=\frac 3 2 (\frac{m}{2\pi\hbar^2})^{\frac 3 2 }\beta^{-\frac 5 2 }
f_{\frac 5 2 }(z_i) (i=1,2),
\end{equation}
where $\beta=1/k_BT,$ $k_B$ is Boltzmann's constant.
$z_1=exp(\beta(\mu-v_0n_2(r)-V_0(r))),$ and
$z_2=exp(\beta(\mu-v_0n_1(r)-V_0(r)))$ are called local
fugacity[19] for component 1 and 2, respectively. $\mu$, the
chemical potential for the atom,
 is determined through $N_i=\int dr n_i(r)$. Before we present the density profile of the
two-component Fermi gases, we consider a range of parameters for
relevant  experiment reported in Ref.[12]. The potentials for the
centre-of-mass motion of a single atom in the hyperfine state can
be approximated as a cylindrically symmetric harmonic potential
with an axial frequency of $\omega_z=2\pi \times 19.5 Hz$ and a
variable radial frequency, which can be varied from
$\omega_r=2\pi\times 44Hz$ to $2\pi\times 370 Hz$. With the
temperature being cooled down, the quantum statistical properties
of the trapped gases become more evident, and at temperature
$T\sim 0.5T_F\sim 300nK,$ the effect of Fermi-Dirac statistical
are observed in the momentum distribution of the gas. With these
parameters, it is evident that $\hbar\omega/k_B T<<1$, i.e., the
semiclassical approximation is a good approach to the realistic
case. (This does not indicate that the semiclassical approximation
holds well at the temperature where BCS-type transition occurs) As
the gas is dilute, we can expand $n_i(r)$ up to first order of the
coupling constant $v_0$, one gives
\begin{equation}
n_a(r)=\frac 3 2 (\frac{m}{2\pi\hbar^2})^{\frac 3 2 }\beta^{-\frac 5 2}(
f_{\frac 5 2 }(z_a^0)-v_0(z_a^0)^2f_{\frac 3 2 }(z_a^0)\beta n_b^0),
\end{equation}
where $$n^0_b(r)=\frac 3 2 (\frac{m}{\pi \hbar^2})^{\frac 3
2}\beta^{-\frac 5 2}f_{\frac 5 2 }(z_b^0),$$ and $a,b =1,2$,
$z_a^0=z_b^0=exp(\beta(\mu-\frac 1 2 m\omega^2 r^2)).$ The result
shows that for $v_0<0$, i.e., attractive interaction, the cloud of
particles is compressed as compared to the noninteracting result.
Because a high density of particles increases the critical
temperature for a BCS-type transition, this effect favors
the formation of the superfluid state[13].\\
\section{Stability properties of a trapped two-component Fermi gas}

Since the experimental realization of the two-component
condensate. Most of the theoretical works concerning
multicomponent condensates[20-26] has been devoted to systems of
two Bose condensates and Bose-Fermi mixture[27-30]. However, other
systems are of fundamental interest,  one of these is
two-component trapped Fermi gas. In fact, by using the sympathetic
cooling,  the fermionic atoms in different hyperfine states has
been cooled down to $300nK$, and the quantum statistical effect in
this system has been reported[12].\\

 For the two-component fermion system, the thermodynamical
properties are trivial if there are not interaction between them.
But in this case the sympathetic cooling scheme can not make
effect and the degenerate fermions in a trapped potential can not
been achieved. The thermodynamical properties may be changed when
the interactions within and between the two hyperfine levels  turn
on. Then a new phenomenon, the phase separation, may occur in this
system. For a homogeneous fermion  mixture system, the Helmholts
free energy can be written as[31]
\begin{eqnarray}
\beta F&=&-\frac{2V}{\lambda_2^3}f_{\frac 5 2 }(z_2)+2a_2N_2\rho_2\lambda_2^2\nonumber\\
&-&
\frac{2V}{\lambda_1^3}f_{\frac 5 2}(z_1)+2 a_1\rho_1N_1\lambda_1^2+
a_{12}(\lambda_1^2+\lambda_2^2)N_2N_1/V,
\end{eqnarray}
where index 2 refers to the fermionic component in level
$|9/2,7/2\rangle$, whereas index 1 stands for them in
$|9/2,9/2\rangle$,  $N_i$ is the number of particles in component
$i$, $\lambda$ denotes the thermal wave length of the atoms,
$f_n(z)$ represents the Fermi integral, $a_i$ and $a_{ij}$ denote
the coupling constants. From eq.(5) we obtain the chemical
potential for each component straightforwardly,
\begin{eqnarray}
\beta\mu_1&=&\beta\mu_1^0+4a_1\rho_1\lambda_1^2+a_{12}(\lambda_1^2+\lambda_2^2)
N_2/V,\nonumber\\
\beta\mu_2&=&\beta\mu_2^0+4a_2\rho_2\lambda_2^2+a_{12}(\lambda_1^2+\lambda_2^2)
N_1/V,
\end{eqnarray}
where $\mu_i^0$ are the chemical potentials of ideal gas. There
are three terms in each chemical potential(6). The second term
comes from the interaction within the component, while the third
term is from the interaction  between the fermions in different
levels. As known, an homogenous binary mixture is stable only when
the symmetric matrix $\hat{\mu}$ given by
\begin{equation}
\hat{\mu}= \left[ \matrix{ \frac{\partial\mu_1}{\partial\rho_1} &
\frac{\partial\mu_1}{\partial\rho_2}\cr
\frac{\partial\mu_2}{\partial\rho_1} &
\frac{\partial\mu_2}{\partial\rho_2} } \right ]
\end{equation}
is non-negatively definite. In other words, all eigenvalues of
matrix $\hat{\mu}$ given in eq.(7)  are non-negative.
Mathematically, for homogenous two-component fermions  the
stability conditions are
\begin{equation}
\frac{\partial \mu_1}{\partial\rho_1}\geq 0,\frac{\partial \mu_2}{\partial\rho_2}\geq0,
\end{equation}
and
\begin{equation}
det\left [ \matrix{ \frac{\partial\mu_1}{\partial\rho_1} &
\frac{\partial\mu_1}{\partial\rho_2}\cr
\frac{\partial\mu_2}{\partial\rho_1} &
\frac{\partial\mu_2}{\partial\rho_2} } \right ] \geq 0.
\end{equation}
For ideal gas, we have $\rho_1=\frac{1}{\lambda_1^3}f_{\frac 3 2
}(z_1),$ and $\rho_2=\frac{1}{\lambda_2^3}f_{\frac 32 }(z_2),$
this leads to
\begin{equation}
\beta\frac{\partial\mu_2^0}{\partial\rho_2}=\frac{\lambda_2^3}{f_{\frac 1 2 }(z_2)},\,
\beta\frac{\partial\mu_2^0}{\partial\rho_2}=\frac{\lambda_1^3}{f_{\frac 1 2}(z_1)},
\end{equation}
It follows from eqs (8) and (9) that
\begin{eqnarray}
&\ &4a_1\lambda_1^2+\frac{\lambda_1^3}{f_{\frac 1 2 }(z_1)}\geq 0,\\
&\ &4a_2\lambda_2^2+\frac{\lambda_2^3}{f_{\frac 1 2 }(z_2)}\geq 0,\\
\mbox{and}\nonumber\\
Z(T,a_{12})&=&(4a_1\lambda_1^2+\frac{\lambda_1^3}{f_{\frac 1 2 }(z_1)})(4a_2\lambda_2^2+
\frac{\lambda_2^3}{f_{\frac 1 2 }(z_2)})\nonumber\\
&-&a_{12}^2(\lambda_1^2+\lambda_2^2)^2
\geq 0.
\end{eqnarray}
Now, we discuss the stability of this system with repulsive
interactions. The case with attractive interactions will be
discussed in the next section. It is obvious that the stability
condition (11) and (12) hold always for $a_1>0$, $a_2>0$. We would
like to point out that the stability conditions (11-13) do not
involve the densities of the both components. At first glance,
this seems to be confusion. In fact, there is no contradiction.
One can demonstrate that at low density the Helmholtz free energy
of the gas reduce to a quadratic form in $N_1$ and $N_2$. To have
a minimum, this form should be positive definite, i.e.,
$\mbox{det}||\frac{\partial^2 F}{\partial N_1\partial
N_2}||\geq0.$ Therefore, the corresponding stability criterion
involves only density-independent constants. This criterion is
similar to the stability conditions for two-component
Bose-Einstein condensate in a trapped ultra-cold gas[23,24,33-37].
When $T\rightarrow \infty, \lambda_i\rightarrow 0$, hence
$Z(T,a_{12})\sim \frac{1}{\rho_1\rho_2}$. Thus at high
temperature, the homogeneous binary gas mixture is always stable
and no phase separation occur. The quantity $Z(T,a_{12})$ as a
function of the temperature is illustrated in Fig.1. We see from
Fig.1  that the system is always stable when $T\rightarrow 0$ and
$T\rightarrow \infty$, and the system is unstable for
$T_{c1}<T<T_{c2}$. In particular, $T_{c1}$ and $T_{c2}$ depend on
$a_{12}$ , the scattering length for fermions in different states.
As $a_{12}$ decreases, $T_{c1}$ tends to $T_{c2}$ (going from
fig.1-a to fig.1-b). \vskip 0cm
/
\begin{figure}

\includegraphics*[width=0.95\columnwidth,
height=0.6\columnwidth]{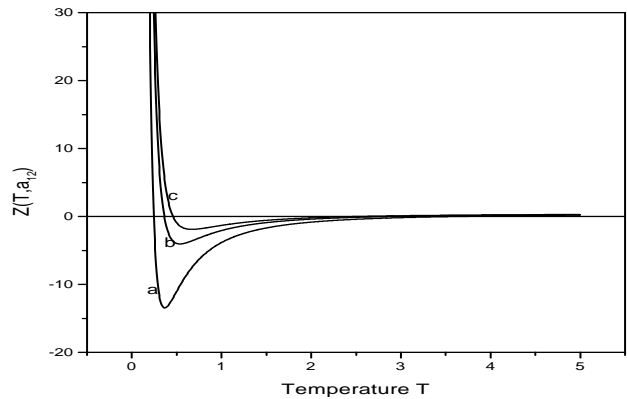} \caption{$Z(T,a_{12})$ vs.
temperature $T$. The coupling constant $a_{12}$ is different for
Fig.1-a and Fig.1-b. a:$a_{12}=0.5$, b:$a_{12}=0.2$. The
parameters for Fig.1c are the same as those in Fig.1-b, but for
the inhomogeneous case.} \label{fig1}
\end{figure}
We would like to note that, the stability condition depends only
on $|a_{12}|$, so the fermions with $a_{12}$ and $-a_{12}$ have
the same stability condition. And the stability discussed above is
only a result of interactions among different components.

Until now, we considered only a homogeneous Fermi gas mixture at
finite temperature. In practise, however, experiments with
ultracold atoms are performed by trapping and cooling atoms in an
external potential that can be generally modeled  by an isotropic
harmonic oscillator $V(r)=\frac m 2\omega^2 r^2$, where $\omega$
is the trapping frequency. An exact criterion for the stability of
an inhomogeneous Fermi-Fermi mixture should involve calculating
the Helmholtz  free energy. Fortunately, in the system considered
here it is a good approach  to take use of the semiclassical
approximation, which treats the atoms as a local homogeneous
system. This approximation requires that the level spacing
$\hbar\omega$ of the trapping potential is much smaller than the
Fermi energy. Of course, the semicalssical approximation always
breaks down at the edge of the gas cloud where the density
vanishes and the effective Fermi energy becomes zero. In this
approximation, the stability conditions can still be calculated by
means discussed above, with the understanding that the effective
chemical potentials are spatially dependent through
$$\mu_1=\mu_1^0-\frac 1 2 m\omega^2r^2,\, \mu_2=\mu_2^0-\frac 1 2
m\omega^2r^2.$$ In this sense, the stability condition is the same
as given in eq.(13) but  replacing $z_i(i=1,2)$ by $$
\tilde{z}_1=z_1e^{-\frac{\beta}{2}m\omega^2r^2},\mbox{\ \ and\ \ }
\tilde{z}_2=z_2e^{-\frac{\beta}{2}m\omega^2 r^2}.$$ As shown in
Fig.1-c, the region of temperature in which the system is unstable
becomes narrow as compared to the homogeneous case($r=0$).
\section{BCS-type transition in trapped two-component Fermi gas}

The achievement of atomic Bose-Einstein condensation has induced
an experimental growth of interest in the properties of ultracold
dilute quantum gases. Of particular interest now is the physics of
trapping and cooling of fermionic atoms. Indeed the prospect of
superfluidity with dilute atomic vapors has already been studied
within the semiclassical approximation by several groups
[13,14,38,39]. Because the semiclassical approximation is not of
fundamental quantum physics,  we will extend the analysis under
the semiclassical approximation by including the discrete nature
of the trapped atom in this section.

Let us consider two species of fermions in a trap, which interact
with each other by two-body collisions( $s-$ wave scattering). The
Hamiltonian describing such a system is given by eq.(1). The two
species of fermions correspond to the trapped atoms in two
hyperfine levels $|1\rangle=|9/2,9/2\rangle$ and
$|2\rangle=|9/2,7/2\rangle$. Expanding
$\psi^{\dagger}_{\alpha}(r)(\alpha=1,2)$ by
\begin{equation}
\psi_{\alpha}^{\dagger}(r)=\sum_n a_{n\alpha}^{\dagger}\phi_n^*(r),
\end{equation}
where $a_{n\alpha}^{\dagger}$ creates one particle in state
$|\alpha n\rangle=|\alpha\rangle\otimes|n\rangle,$ which satisfies
\begin{equation}
(-\frac{\hbar^2}{2m}\nabla^2+\frac 1 2 m\omega^2r^2)|\alpha n\rangle=
\varepsilon_{n\alpha}|\alpha n\rangle,
\end{equation}
the Hamiltonian in eq.(1) becomes
\begin{equation}
H=\sum_{\alpha n}\varepsilon_{\alpha n}a_{\alpha
n}^{\dagger}a_{\alpha n}+
\sum_{i,j,m,n}V_{ijmn}a_{i\alpha}^{\dagger}a_{j-\alpha}^{\dagger}a_{m-\alpha}a_{n\alpha},
\end{equation}
where $V_{ijmn}=v_0\int dr
dr^{'}\delta(r-r^{'})\phi_i^*(r)\phi_j^*(r^{'})\phi_m(r^{'})\phi_n(r)$
and $\alpha$, $-\alpha$ denote the two hyperfine levels(say, for
example, if $\alpha=|9/2,9/2\rangle$ , then
$-\alpha=|9/2,7/2\rangle$). The next step in a mean-field
treatment of the Hamiltonian in eq.(16) is to develop the operator
product $\psi_{\alpha n}^{\dagger}\psi_{-\alpha n}^{\dagger}$
around their mean values by substituting $$\psi_{\alpha
n}^{\dagger} \psi_{-\alpha n}^{\dagger}=\langle \psi_{\alpha
n}^{\dagger}\psi_{-\alpha n}^{\dagger}\rangle +\delta \psi_{\alpha
n}^{\dagger}\psi_{-\alpha n}^{\dagger}.$$ To first order in the
fluctuations, we are left with the effective mean-field
Hamiltonian
\begin{equation}
H\simeq \sum_{\alpha n}\varepsilon_{\alpha n}a_{\alpha
n}^{\dagger}a_{\alpha n} +\sum_m \Delta_m a_{\alpha m}^{\dagger}
a_{-\alpha m}+\sum_m \Delta^*_m a_{\alpha m}a_{-\alpha m}.
\end{equation}
Here $\Delta_m=(\Delta_m^*)^*=\sum_nV_{mmnn}\langle a_{\alpha
n}a_{-\alpha n}\rangle$ is the equilibrium value of the BCS order
parameter. As the effective mean-field Hamiltonian in terms of the
operators $a_{\alpha n}^{\dagger}$ and $a_{\alpha n}$ is
non-diagonal, one can not directly calculate the expection value
$\langle a_{\alpha n}^{\dagger}a_{\alpha n}\rangle$. This is, as
usual, resolved by first applying a Bogoliubov transformation
 \begin{eqnarray}
b_{\alpha m}=u_ma_{\alpha m}+v_m a_{-\alpha m}^{\dagger},\nonumber\\
b_{\alpha m}^{\dagger}=u_m a_{\alpha m}^{\dagger}+v_m a_{-\alpha m},
\end{eqnarray}
to diagonalize the Hamiltonian in eq.(17). After performing this
unitary transformation, we require that the Hamiltonian in terms
of the new quasiparticle operators $b_{\alpha m}$ and $b_{\alpha
m}^{\dagger}$
 has only diagonal elements, and furthermore  we assume that these operators
 obey the usual
 anticommutation
relations still. This determines the values of the yet unknown
$u_m$ and $v_m$. The latter constraint requires that the constant
$u_m$ and $v_m$ must satisfies the relations $|u_m|^2+|v_m|^2=1$
and the requirement of diagonality of the Hamiltonian after
Bogoluibov transformation lead to
$|v_m|^2=\frac{2}{1-\varepsilon_m/E_m}$ with $\varepsilon_m=
\varepsilon_{\alpha m},$ $E_m=\sqrt{\Delta_m^2+\varepsilon_m^2}$.
$ E_m$ are eigenvalues of the Bogoluibov quasiparticles.

Using Bogoluibov transformation(18), the equilibrium value of the
BCS order parameter is calculated easily, it is given that
\begin{equation}
\Delta_n=\sum_mV_{mmnn}\frac{\Delta_m}{E_m}tanh\frac{\beta}{2}E_m.
\end{equation}
As usual, the order parameter does not depend on its index $m$. So
we arrive at the gap equation
\begin{equation}
\sum_m
V_{mmnn}\frac{1}{\sqrt{\varepsilon_m^2+\Delta^2}}tanh\frac{\sqrt{\varepsilon_m^2+
\Delta^2}}{2k_B T}=1.
\end{equation}
Setting $\Delta=0$, one finds the critical temperature $T_c$ as a
function of the trapping frequency $\omega$[18]. The density
$N(0)$ of atoms near the Fermi surface and the coupling constant
$v_0=V_{mmnn}$,
\begin{eqnarray}
1&=&v_0N(0) ln(1.13\beta_c\hbar\omega)+v_0Bk_BT_c\frac{exp[\frac{\hbar\omega}{k_BT_c}]}{1+
exp[\frac{\hbar\omega}{k_BT_c}]}\nonumber\\
&-&\frac 1 2 v_0Bk_BT_c-
v_0Bk_BT_cln(1+exp[\frac{\hbar\omega}{k_BT_c}])\nonumber\\
&+&v_0Bk_BT_cln2,
\end{eqnarray}
where $B=\frac {3}{2(\hbar\omega)^2}$ and
$\beta_c=\frac{1}{k_BT_c}.$ The first term in the right hand side
of eq.(21) is just from the usual BCS theory, in other words, if
the semiclassical approximation is a good approach to the theory
or the discrete nature of the trap levels can be neglected,
$T_c=\frac{\hbar\omega}{k_B}e^{-\frac{1}{v_0N(0)}}.$ The rest
terms in the right hand side of eq.(21) are corrections of the
discrete trap levels to the usual BCS theory.

This superfluid phase transition, which is similar to the BCS
transition in a superconductor, might occur at very low
temperature. At such low temperature, whether the semiclassical
approximation holds depend on both the temperature and the
trapping frequency. Hence we investigate here the BCS-type
transition from the other aspect, beyond the semiclassical
approximation. The results show that if the semiclassical
approximation holds, i.e., $\hbar\omega/k_BT_c<<1$, the transition
temperature is just the BCS one. Otherwise the effects of the
discrete trap levels provide a negative correction to the
transition temperature.\\
\section{Conclusion}
In summary, we considered a Fermi gas occupying two hyperfine
states trapped in a magnetic field. Atoms in different hyperfine
levels can interact via $s-$wave scattering. Under the
semiclassical approximation, we calculated  and discussed the
density profile for the trapped fermions. The purterbative results
up to the first order of the coupling constant show that the
density of the atom is compressed(diluted) as compared to the
noninteracting case due to attractive (repulsive)interaction. We
also investigate the mechanical and statistical stability of the
two-component gas with interaction between the atoms in different
hyperfine levels, and find  that these interactions strongly
affect the stability of the system at finite temperature. The
regime $T_{c1}\leq T\leq T_{c2}$ in which the system is unstable
depend on the strength of the interaction and the spatial atomic
position. Furthermore, we consider the BCS-type transition beyond
the semiclassical approximation, within which the most current
literature study the superfluid state of trapped fermonic atoms.
The results showed  the transition temperature is a function of
the trapped frequency, the coupling constant and, as usual, the
density of the atoms near the Fermi surface. Neglecting the
discrete nature of the trap levels, the transition temperature
 return to the BCS's one, and the correction to $T_c$ is negative due
 to the discrete nature of the trapped
atom.
\\
{\bf \large ACKNOWLEDGEMENT:}\\ We thank Prof. C. P. Sun, Prof. W.
M. Zheng,  Dr. Li You for their stimulating discussions. This work was supported by
NNSF of China.\\

\end{document}